# Soft Modes of Collective Domain-Wall Vibrations in Epitaxial Ferroelectric Thin Films

N. A. Pertsev and A. Yu. Emelyanov

*A. F. Ioffe Physico-Technical Institute, Russian Academy of Sciences, 194021 St. Petersburg, Russia*

Mechanical restoring forces acting on ferroelastic domain walls displaced from the equilibrium positions in epitaxial films are calculated for various modes of their cooperative translational oscillations. For vibrations of the domain-wall superlattice with the wave vectors corresponding to the center and boundaries of the first Brillouin zone, the soft modes are singled out that are distinguished by a minimum magnitude of the restoring force. It is shown that, in polydomain ferroelectric thin films, the soft modes of wall vibrations may create enormously large contribution to the film permittivity.

The formation of multiple patterns of elastic domains (twins) is a characteristic feature of epitaxial ferroelectric thin films [1–6]. The experimental data [3,7] and theoretical considerations [8,9] suggest that in perovskite ferroelectrics the $90^0$ domain walls (DWs) may be highly mobile on the microscopic scale even at room temperature [10]. Accordingly, the forced translational vibrations of $90^0$ walls are believed to contribute considerably to the small-signal dielectric and piezoelectric responses of ferroelectric thin films [9,11]. Up to know, however, only one specific mode of DW vibrations in epitaxial films, namely the antiparallel motion of neighboring walls, was described theoretically [9,11].

In this Letter, a general analysis of cooperative translational vibrations of ferroelastic walls in epitaxial films is carried out. The calculations are performed in a conventional linear elastic approximation and based on the dislocation-disclination modeling of the sources of internal stresses in polydomain films [12,13]. Forced low-frequency vibrations of $90^0$ walls in epitaxial thin films with a laminar *c*/*a*/*c*/*a* structure are considered, and soft modes of DW vibrations are singled out, which are characterized by enhanced collective mobility of domain walls. It is shown that the excitation of a soft mode in a polydomain ferroelectric film may increase drastically the film dielectric response.

Consider a laminar $90^0$ domain structure with the walls inclined at $45^0$ to the film/substrate interface Σ (Fig. 1). Domain patterns of this type are widely observed in epitaxial films of perovskite ferroelectrics grown on (001)-oriented cubic substrates [3–5]. They consist of alternating elastic domains with the polar *c* axis orthogonal and parallel to Σ (*c* and *a* domains). Following Refs. 1,8, we shall assume that in equilibrium the *c*/*a*/*c*/*a* structure has an exact periodcity. Then the initial domain geometry may be described by the width *d* of the *c* domains and the domain period *D*, both measured along the interface Σ. Since *c*/*a* and *a*/*c* walls create disclinations of opposite sign at their junctions with Σ [12,13], they should be considered as *physically distinct*. (This approach makes our theory essentially different from the description of twin-wall vibrations in bulk ferroelastic crystals, which was developed earlier in Refs. 14–15.) Therefore, the domain pattern should be regarded as a superposition of two periodic arrays of equivalent DWs, shifted by the distance *d* from each other. Accordingly, displacements of the *c*/*a* and *a*/*c* walls from their initial equilibrium positions will be denoted here as $\delta_m^{(1)}$ and $\delta_n^{(2)}$, respectively, where the integer numbers *m* and *n* define positions of individual walls in the film. Displacements $\delta_m^{(1)}$ and $\delta_n^{(2)}$ will be considered as small quantities in comparison with the initial domain sizes *d* and $D - d$.

Let us calculate now the variation $\Delta U$ of the energy of a polydomain film/substrate system, which is caused by translational vibrations of $90^0$ walls. Restricting our analysis by the low-frequency range $\Omega << c_t/H$ ($c_t$ is the velocity of transverse sound wave, $H$ is the film thickness), we can neglect the kinetic energy of the medium [16] and evaluate $\Delta U$ in a conventional quasi-static linear elastic approximation [8]. Then in the absence of external fields the sought variation will be reduced to the change $\Delta W$ of the elastic energy and can be represented as the following quadratic form:

$$\Delta W = \sum_{m,n} \left\{ \frac{1}{2}\chi\left[(m-n)D\right](\delta_m^{(1)}\delta_n^{(1)} + \delta_m^{(2)}\delta_n^{(2)}) - \chi\left[d + (m-n)D\right]\delta_m^{(1)}\delta_n^{(2)} \right\}. \quad (1)$$

In deriving Eq. (1) we have taken into account that the elastic interaction between *m*-th and *n*-th walls depends only on the distance *l* between them, but not on their positions in the film. The negative sign of the second term in curly brackets in Eq. (1) reflects the fact that displacements of $c/a$ and $a/c$ walls in the same direction result in opposite changes of the sources of internal stresses in the epitaxial system. As shown in Ref. 11, these changes are equivalent to the introduction of additional edge dislocations at the junctions of domain walls with the interface. Dislocations with the Burgers vectors perpendicular to $\Sigma$ allow for shifts of the wedge disclinations modeling the junctions of $c/a$ and $a/c$ walls with $\Sigma$ (see Fig. 1). In turn, the appearance of dislocations with the Burgers vectors parallel to $\Sigma$ is due to changes in the distribution of dislocations modeling the interface, which take place on the sections of $\Sigma$ swept by moving walls [11].

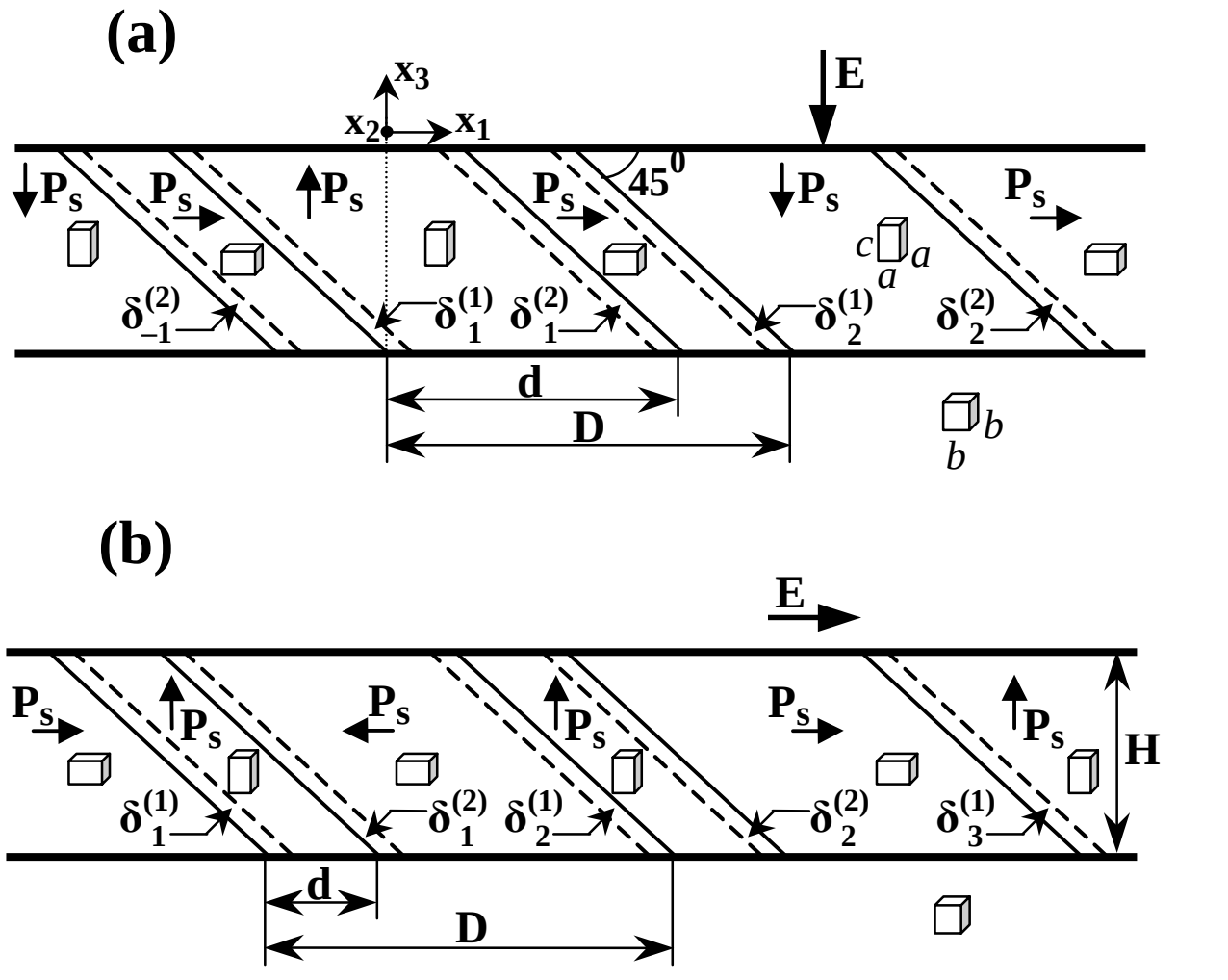

FIG. 1. Soft $c$-mode (a) and $a$-mode (b) of collective vibrations of $90^0$ walls in ferroelectric films with the laminar $c/a/c/a$ structure and special polarization distributions. DW shifts in external electric field **E** are shown by dashed lines.

The function $\chi(l)$ involved in Eq. (1) can be derived via the calculation of the energy of elastic interaction between parallel edge dislocations situated at the distance $H$ from the boundary of elastic half-space. This energy may be found by integrating the stresses $T_{ij}$ of one dislocation over the surface of a cut necessary for the formation of another dislocation. To calculate the stress field of an edge dislocation near a free surface, it is convenient to represent it as a dipole of wedge disclinations with an infinitesimal arm [17]. Assuming the film/substrate system to be elastically homogeneous and isotropic with an effective shear modulus $G$ and Poisson′s ratio $\nu$, we can find $T_{ij}$ by differentiating the disclination stress field in elastic half-space, which is given in Ref. 17. The integration of $T_{ij}$ yields the following expression for the function $\chi(l)$ calculated per unit DW length along the $x_2$ axis:

$$\chi(l) = \frac{G(S_c - S_a)^2}{\pi(1-\nu)}\left[\ln\left(1+\frac{4H^2}{l^2}\right) + \frac{4H^2(l^2-4H^2)}{(l^2+4H^2)^2}\right], \quad (2)$$

where $S_a = (b^* - a)/a$ and $S_c = (b^* - c)/c$ are the misfit strains between the film (assumed to have a tetragonal lattice with the parameters $a$ and $c > a$ in the free state) and a cubic substrate with the effective lattice parameter $b^*$ [18]. We have taken into account that the dislocation Burgers vectors are proportional to DW displacements [11], with $\pm(S_a - S_c)$ being the proportionality constants.

Equation (2) defines coefficients of all terms in the quadratic form (1), except for the factor $\chi(0)$ that characterizes the self-action of the displaced wall. Indeed, at $l \to 0$ the function $\chi(l)$ diverges logarithmically so that our method cannot be used to calculate $\chi(0)$. However, this factor may be found from the condition of the elastic-energy invariance with respect to the translation of DW "superlattice" as a whole. Assuming displacements $\delta_m^{(1)}$ and $\delta_n^{(2)}$ of all DWs to be the same, from the condition $\Delta W = 0$ we obtain

$$\sum_{m,n}\left\{\chi\left[(m-n)D\right] - \chi\left[d+(m-n)D\right]\right\} = 0 \quad (3)$$

so that ($k = \pm1, \pm2, \ldots$)

$$\chi(0) = \chi(d) - \sum_k\left[\ \chi(kD) - \chi(d+kD)\ \right]. \quad (4)$$

Let us represent now the displacements of $c/a$ and $a/c$ walls in the form of Fourier series:

$$\delta_n^{(1)} = \sum_q \hat{\delta}^{(1)}(q)e^{iqnD}, \quad \delta_n^{(2)} = \sum_q \hat{\delta}^{(2)}(q)e^{iqnD}. \quad (5)$$

The Fourier transforms $\hat{\delta}^{(1)}(q)$ and $\hat{\delta}^{(2)}(q)$ of the wall displacements are defined by the relations

$$\hat{\delta}^{(1)}(q) = \lim_{N\to\infty}\frac{1}{N}\sum_n \delta_n^{(1)}e^{-iqnD}, \quad \hat{\delta}^{(2)}(q) = \lim_{N\to\infty}\frac{1}{N}\sum_n \delta_n^{(2)}e^{-iqnD} \quad (6)$$

where $N$ is the total number of DWs in an array, over which the summation is carried out. Using Eqs. (5), which represent displacements of individual walls as a superposition of the DW-lattice vibrations having various wave vectors $q$, we can rewrite the quadratic form (1) as

$$\Delta w = \frac{1}{2}\sum_q \hat{\chi}_0(q)\left[\hat{\delta}^{(1)}(q)\hat{\delta}^{(1)}(-q) + \hat{\delta}^{(2)}(q)\hat{\delta}^{(2)}(-q)\right] - \sum_q \hat{\chi}_d(q)\hat{\delta}^{(1)}(q)\hat{\delta}^{(2)}(-q), \quad (7)$$

where $\Delta w$ is the change of the elastic energy $\Delta W$ calculated per "unit cell" of the domain-wall superlattice (containing two neighboring walls), and the following designations were introduced:

$$\hat{\chi}_0(q) = \sum_n \chi(nD)\, e^{-iqnD} \ ,$$
$$\hat{\chi}_d(q) = \sum_n \chi(d+nD)\, e^{-iqnD} \, . \qquad (8)$$

Equation (7) makes it possible to compare the mobilities of domain walls at different modes of their collective vibrations. In this paper we shall consider only modes with the wave vectors $q = 0$ and $q = \pm\,\pi/D$, which correspond to the center and boundaries of the first Brillouin zone. In this case, $\hat{\delta}^{(1)}(q)$ and $\hat{\delta}^{(2)}(q)$ are real numbers so that the energy $\Delta w$ becomes a function of only two variables at a fixed wave vector $q$. It is useful to introduce now the new collective coordinates $r(q)$ and $\varphi(q)$ satisfying the relations $\hat{\delta}^{(1)}(q) = r(q)\cos\varphi(q)$ and $\hat{\delta}^{(2)}(q) = r(q)\sin\varphi(q)$. Evidently, the "radius" $r(q)$ defines the deviation of the DW superlattice as a whole from the equilibrium state at the appearance of a mode with the wave vector $q$. The "polar angle" $\varphi(q)$ shows to which extent the *c*/*a* and *a*/*c* walls are involved in the vibrations: at $\varphi(q) = 0$ or $\varphi(q) = \pi/2$ only the *c*/*a* or *a*/*c* walls oscillate, whereas at $\varphi(q) = \pi/4$ the walls of both types participate equally in the vibrations.

The mechanical restoring force, which hinders the development of a certain mode of cooperative DW vibrations, is defined by the derivative of the energy $\Delta w$ with respect to the coordinate $r(q)$. Using Eq. (7) and taking into account that at $q = 0$ and $q = \pm\,\pi/D$ the equalities $\hat{\delta}^{(1)}(q) = \hat{\delta}^{(1)}(-q)$ and $\hat{\delta}^{(2)}(q) = \hat{\delta}^{(2)}(-q)$ hold, we obtain

$$\hat{f}(q) = -\frac{\partial \Delta w}{\partial r(q)} = -r(q)\left[\hat{\chi}_0(q) - \hat{\chi}_d(q)\sin 2\varphi(q)\right]. \qquad (9)$$

The magnitude of $\hat{f}(q)$ depends on the parameter $\varphi(q)$ and reaches extreme values provided that

$$\frac{\partial \hat{f}(q)}{\partial \varphi(q)} = 2r(q)\, \hat{\chi}_d(q)\cos 2\varphi(q) = 0\, . \qquad (10)$$

The condition (10) gives $\varphi(q) = \pi/4 + \pi k/2$ with $k = \pm 1, \pm 2, \ldots$ Therefore, the maximum and minimum DW mobilities correspond to those vibration modes, which are characterized by the relationship $\hat{\delta}^{(1)}(q) = \hat{\delta}^{(2)}(q)$ or $\hat{\delta}^{(1)}(q) = -\hat{\delta}^{(2)}(q)$ between the Fourier transforms of the displacements of *c*/*a* and *a*/*c* walls.

Using Eq. (5), it can be easily shown that at $q = 0$ the modes with extreme magnitudes of the mechanical restoring force turn out to be the rigid translation of the DW superlattice ($\delta_n^{(1)} = \delta_n^{(2)} = \delta$) and the antiparallel motion of *c*/*a* and *a*/*c* walls ($\delta_n^{(1)} = -\delta_n^{(2)} = \delta$), which was discussed in Ref. 8. In turn, at $q = \pm\pi/D$ the relations defining two "extreme" modes take the form

$$\delta_n^{(1)} = -\delta_n^{(2)} = (-1)^n \delta\, . \qquad (11)$$

$$\delta_n^{(1)} = \delta_n^{(2)} = (-1)^n \delta\, , \qquad (12)$$

The perturbation of DW superlattice described by Eq. (11) represents the antiparallel motion of neighboring *a* domains, during which their sizes remain constant (Fig. 1*a*). The widths of neighboring *c* domains change in the opposite way so that vibrations of this type may be termed *c-mode*. Equation (12) describes a type of motion pairwise to the *c*-mode (*a-mode*), where the widths of neighboring *a* domains change in opposite phase, whereas the *c* domains shift as a whole (Fig. 1*b*).

For revealed extreme modes, it is necessary to calculate now the magnitude of the restoring force $f_{res} = -(1/2\sqrt{2}H)\, \partial\Delta w/\partial\delta$ acting per unit area of a displaced wall. At $q = 0$ Eq. (7) gives $\Delta w = [\, \hat{\chi}_0(0) \mp \hat{\chi}_d(0)\, ]\delta^2$, where the upper sign refers to the rigid translation of DW superlattice, whereas the lower one – to the antiparallel motion of *c*/*a* and *a*/*c* walls (*h*-mode). From Eqs. (8) and (4) it follows that $\hat{\chi}_0(0) = \hat{\chi}_d(0)$. Accordingly, in the case of the rigid translation the restoring force $f_{res} = -\,k\,\delta$ goes to zero in agreement with our initial supposition. For the *h*-mode, this force is defined by constant $k_h = (\sqrt{2}/H)\hat{\chi}_d(0)$. Substituting Eq. (2) into Eq. (8) and summing up the series at $q$=0, one can calculate $\hat{\chi}_d(0)$ and obtain the following expression for the constant $k_h$:

$$k_h = \frac{\sqrt{2}G(S_a - S_c)^2}{\pi(1-\nu)H}\left[R\left(\frac{d}{2D}, \frac{H}{2D}\right) + R\left(\frac{d}{2D} + \frac{1}{2}, \frac{H}{2D}\right)\right], \qquad (13)$$

where the function $R(x, y)$ is given by

$$R(x,y) = \ln\left[\frac{\cosh(4\pi y) - \cos(2\pi x)}{1 - \cos(2\pi x)}\right] - 8\pi^2 y^2 \frac{[\cosh(4\pi y)\cos(2\pi x) - 1]}{[\cosh(4\pi y) - \cos(2\pi x)]^2}\, . \qquad (14)$$

It should be noted that earlier [8] the force constant $k_h$ was derived as $k_h = [\sqrt{2}\,G(S_a - S_c)^2 / \pi(1-\nu)H]\; R(d/D, H/D)$ with the function $R\,(x, y)$ also given by Eq. (14). The mathematical analysis shows that Eq. (13) can be cast into this form as well, which supports the validity of our results. Numerical calculations also demonstrate that $k_h$ is positive at all possible values of the parameters $H/D$ and $0 < d/D < 1$ of the domain pattern.

In turn, at $q = \pm\,\pi/D$ the elastic-energy change $\Delta w$ caused by the appearance of an extreme mode is given by $\Delta w = [\,\hat{\chi}_0(\pi/D) \pm \hat{\chi}_d(\pi/D)\,]\delta^2$, where the upper and lower signs refer to the *c*- and *a*-mode of collective DW vibrations, respectively. By summing the series (8) at $q = \pm\,\pi/D$ with the account of Eqs. (2) and (4), it is possible to find $\hat{\chi}_0(\pi/D)$ and $\hat{\chi}_d(\pi/D)$. The substitution of the results in the above formula yields the following expressions for the constants $k_{sc}$ and $k_{sa}$ defining restoring forces for the *c*- and *a*-modes:

$$k_{sc} = \frac{\sqrt{2}G(S_a - S_c)^2}{\pi(1-\nu)H}\left[R\left(\frac{d}{2D}, \frac{H}{2D}\right) - R\left(\frac{1}{2}, \frac{H}{2D}\right)\right]. \quad (15)$$

$$k_{sa} = \frac{\sqrt{2}G(S_a - S_c)^2}{\pi(1-\nu)H}\left[R\left(\frac{d}{2D} + \frac{1}{2}, \frac{H}{2D}\right) - R\left(\frac{1}{2}, \frac{H}{2D}\right)\right], \quad (16)$$

Numerical calculations based on Eqs. (15)-(16) show that the constants $k_{sc}$ and $k_{sa}$ are positive at all allowed values of $H/D$ and $d$/D. This result indicates the stability of the initial *c*/*a*/*c*/*a* structure.

Compare now the constants $k_h$, $k_{sc}$, and $k_{sa}$. The analysis of Eqs. (13)-(16) demonstrate that $k_h$ is considerably larger than $k_{sc}$ and $k_{sa}$ at any geometry of the *c*/*a*/*c*/*a* structure. This feature can be explained by the fact that the *h*-mode changes the fraction $\phi_c = V_c / V$ of the film volume occupied by *c* domains, whereas the *c*- and *a*-modes leave $\phi_c$ unaltered (see Fig. 1). Therefore, the antiparallel motion of *c*/*a* and *a*/*c* walls appears to be the most *hard mode* of collective DW vibrations.

In turn, the force constant $k_{sc}$ at $d/D > 0.5$ appears to be smaller than $k_{sa}$, whereas at $d/D <$ 0.5 – larger than $k_{sa}$. For the wave vector $q = \pi/D$, therefore, the minimum restoring force corresponds to the *c*-mode at $d/D > 0.5$ and to the *a*-mode at $d/D < 0.5$. Accordingly, in the domain structures, where the equilibrium width of *c* domains is larger than that of *a* domains, the *c*-mode may be regarded as a *soft mode* of DW vibrations. When the relation between the sizes of these domains is opposite ($d/D < 0.5$), the *a*-mode becomes the soft mode of these oscillations. In the special case of $d/D = 0.5$, the equality $k_{sc} = k_{sa}$ holds, and all the modes with $q = \pi/D$ have the same restoring force because in Eq. (9) $\hat{\chi}_d(\pi/D) = 0$ at $d/D = 0.5$.

Equations (13), (15), and (16) show that the force constants given in units of $G\,(S_a - S_c)^2 / [(1-\nu)H\,]$ depend on the parameters $D/H$ and $d/D$ of unperturbed domain pattern only. For the constant $k_{sc}$, these dependences are shown in Fig. 2. It can be seen that $k_{sc}$ decreases gradually with increasing volume fraction $\phi_c = d/D$ of *c* domains in the initial structure, but it is a nonmonotonic function of the normalized domain period $D/H$. The constant $k_h$, which characterizes the hard mode, depends on $\phi_c$ and $D/H$ in a very different way. In particular, when the domain structure becomes dense ($D/H$ << 1), the constant $k_{sc}$ tends to a finite limit at $D/H \to 0$, whereas $k_h \sim D^{-1}$ increases continuously [8].

Let us analyze now how the extreme modes may be excited during the forced vibrations of $90^0$ walls in epitaxial films. In ferroelectric films, translational DW vibrations can be induced by an external ac electric field $E(t) = E_m \sin\Omega t$. In a conventional plate-capacitor setup, where the film is sandwiched between two continuous electrodes, the field **E** interacts mainly with the spontaneous polarization $\mathbf{P}_s$ in the *c* domains since $\mathbf{P}_s$ is orthogonal to the film surfaces here. When in all *c* domains the vector $\mathbf{P}_s$ has the same orientation, the measuring field $E(t)$ induces the antiparallel motion of *c*/*a* and *a*/*c* walls (*h*-mode) [8]. This situation takes place in the films polarized by a strong dc electric field exceeding the critical field necessary for the $180^0$ switching.

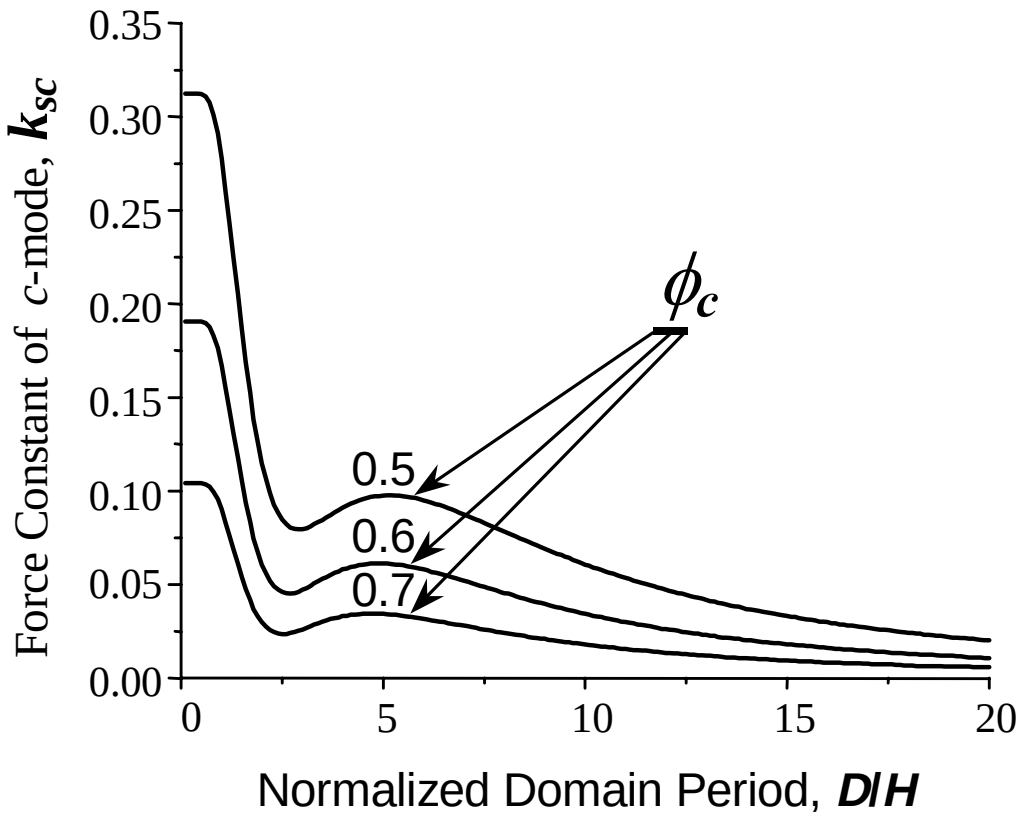

FIG. 2. Normalized force constant $k_{sc}\,(1-\nu)H\,/G(S_c - S_a)^2$ of the *c*-mode as a function of the domain-wall periodicity $D/H$ at three different values of $\phi_c = d/D$. Since according to Eqs. (15)-(16) $k_{sc}(1-\phi_c) = k_{sa}(\phi_c)$, these curves can be used to describe the constant $k_{sa}$ of the *a*-mode as well.

The excitation of the soft $c$-mode appears to be much more complicated technical problem. To achieve this goal, it is necessary to create a spatially inhomogeneous electric field $\mathbf{E}(x_1)$ in the epitaxial layer. The component $E_3$ of this field, which is orthogonal to the film surfaces, must be modulated in the film plane so as to be opposite in sign in the neighboring $c$ domains at the same time. Such a distribution of external field cannot be created in setups with continuous electrodes. Instead, fingered electrodes with a special geometry, which provides the field modulation along the $x_1$ axis fitting the DW periodicity $D$, must be deposited on the substrate boundary and the film free surface.

The system of fingered electrodes may be used in two different ways. First, the $c$-mode may be induced in a prepolarized film with the usual spatial distribution of $\mathbf{P}_s$, where the vector $\mathbf{P}_s$ has the same orientation in all $c$ domains. In this case, strong dc field with a fixed polarity must be applied to the film during its cooling through the Curie temperature in order to promote formation of $c$ domains under the electrodes and to create uniform polarizations $\mathbf{P}_s$ inside them. After this preliminary poling of the film, a weak ac field $E_3(t)$ should be induced between the upper and lower electrodes in such a way that it has opposite polarities for the neighboring pairs of electrodes at any moment. (Then one of two neighboring $c$ domains will shrink simultaneously with the widening of the other.) Second, this system of electrodes may be used to construct a special initial distribution of the spontaneous polarization $\mathbf{P}_s$ in the film, where $\mathbf{P}_s$ has opposite orientations in the neighboring $c$ domains (Fig. 1$a$). The application of ac field in the same phase to all pairs of electrodes will induce the soft $c$-mode in this film, instead of the hard $h$-mode appearing in conventionally polarized films.

In the case of dielectric measurements performed on a film with the above special distribution of polarization, the DW contribution to the film permittivity $\varepsilon_{33}$ will be caused by the $c$-mode of collective wall vibrations. This contribution $\Delta\varepsilon_{33}$ may be calculated in the same way as it was done for the $h$-mode in Ref. 8. The driving force $f_{\text{driv}}$ acting on a wall in a weak measuring field $E \sim 1$ kV/m can be evaluated as $f_{\text{driv}} = P_s E_3(t) = P_s E_m \sin\Omega t$ per unit DW area. For fields with comparatively low frequencies $\Omega << c_t/H$, the wall displacement $\delta(t)$ at the moment $t$ is defined simply by the force equilibrium $f_{\text{driv}}(t) + f_{\text{res}}(t) = 0$. Since in the quasi-static approximation the restoring force equals $f_{\text{res}}(t) = -k_{sc}\,\delta(t)$, we find the wall displacement as $\delta(t) = P_s E_3(t)/k_{sc}$. Collective DW movements in the form of the soft $c$-mode are accompanied by the variation $\Delta P_3(t) = 2\sqrt{2}\,P_s\,\delta(t)/D$ of the mean polarization in a film. Accordingly, a DW contribution $\Delta\varepsilon_{33}$ to the film dielectric response appears, given by formula ($\varepsilon_0$ is the permittivity of the vacuum)

$$\Delta\varepsilon_{33} = \frac{\Delta P_3(t)}{\varepsilon_0 E_3(t)} = \frac{2\sqrt{2}\,P_s^2}{\varepsilon_0\,k_{sc}\,D}\,. \qquad (17)$$

The substitution of Eq. (15) for the force constant $k_{sc}$ into Eq. (17) shows that $\Delta\varepsilon_{33}$ can be represented as a product of the material parameter $\eta = P_s^2(1-\nu)/\,\varepsilon_0 G(S_a - S_c)^2$ and a dimensionless function of the normalized domain period $D/H$ and the domain population $\phi_c = d/D$. Since the equilibrium values of $D/H$ and $\phi_c$ are known functions of the normalized film thickness $H/H_0$ and the relative coherency strain $S_r = (b^* - a)/(c - a)$ in the epitaxy [13], we can calculate the dependences of the DW contribution $\Delta\varepsilon_{33}$ on these parameters. Some of our results are shown in Fig. 3.

The dependence of $\Delta\varepsilon_{33}$ on the film thickness $H$ is marked by the presence of a local maximum and by the increase of $\Delta\varepsilon_{33}$ at $H \to 0$ and $H \to \infty$. The growth of DW contribution in thick films is caused by the fact that at $H >> 100H_0$, when the domain pattern becomes dense ($D << H$), the force constant $k_{sc}$ decreases as $H^{-1}$. In this case Eq. (17) gives $\Delta\varepsilon_{33} \sim H/D$. Accordingly, the DW contribution $\Delta\varepsilon_{33}$ increases proportionally to $H^{1/2}$ at $H \to \infty$, since at $D << H$ the period $D \sim H^{1/2}$ [13]. This feature distinguishes the $c$-mode from the hard $h$-mode giving a contribution $\Delta\varepsilon_{33}$, which tends to a finite limit of $\Delta\varepsilon_{33} = 0.5\eta$ at $H \to \infty$ [8].

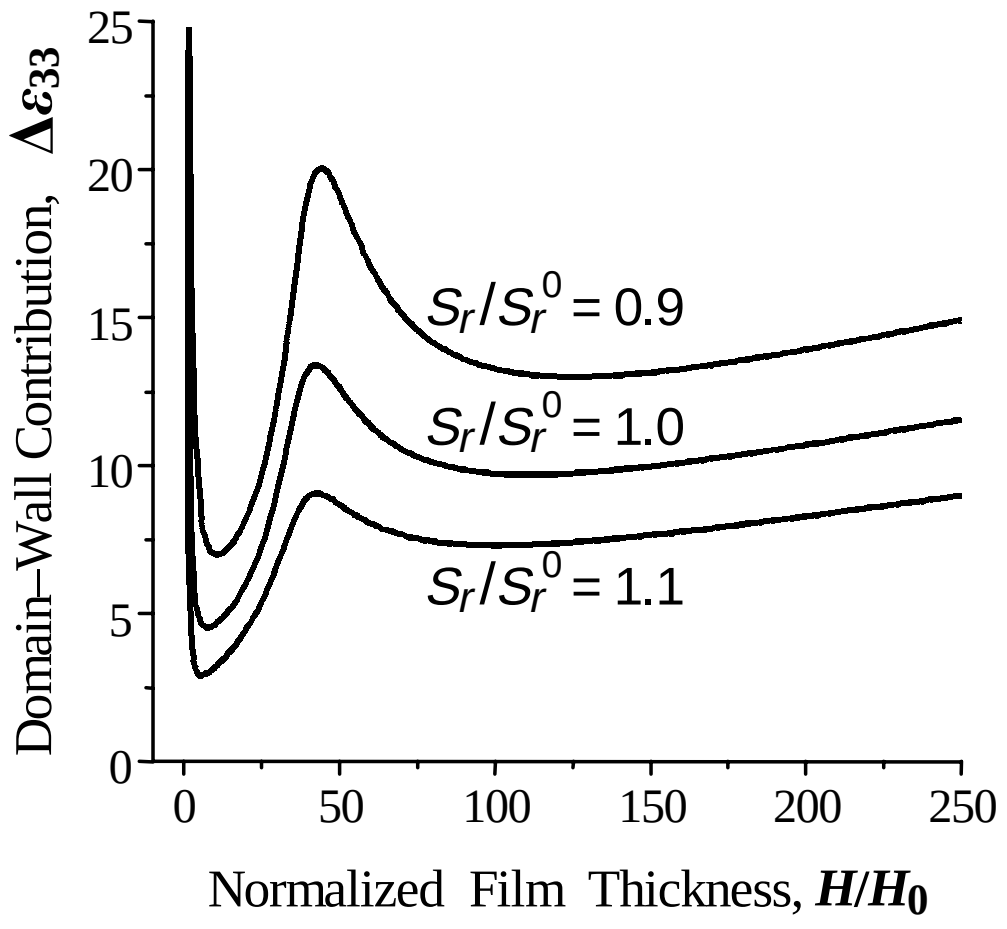

FIG. 3. Normalized $c$-mode contribution $\Delta\varepsilon_{33}\,\varepsilon_0 G(S_a - S_c)^2\,/\,P_s^2(1-\nu)$ to the dielectric permittivity as a function of the film thickness $H/H_0$. The curves are plotted at different values of $S_r/S_r^0$, where $S_r^0 = 1/\,2(1+\nu)$.

The most important feature, however, consists in the fact that the *c*-mode creates enormously large contribution to the permittivity of ferroelectric films with a conventional thickness of $H$ = (0.1–1) μm. Indeed, at $H >> H_0$ ($H_0$ ~1 nm for $BaTiO_3$ and $PbTiO_3$ [8,9]) the *h*-mode gives only $\Delta\varepsilon_{33} \sim 0.5\eta$, whereas the *c*-mode provides $\Delta\varepsilon_{33} \sim 10\eta$ and even more in a considerable range of misfit strains $S_r$. Accordingly, in $BaTiO_3$ and $Pb(Zr_{0.51}Ti_{0.49})O_3$ films, for example, the contribution $\Delta\varepsilon_{33}$ of the *c*-mode at room temperature appears to be larger than 20000, since in this case $\eta \approx 2000$ [8].

For the excitation of the *a*-mode in polydomain ferroelectric films, it is necessary to create a spatially inhomogeneous electric field directed in the film plane ($E_1 \neq 0$). Theoretically, the soft *a*-mode may give anomalous contribution to the film in-plane permittivity $\varepsilon_{11}$.

Thus, in polydomain ferroelectric thin films, the measuring electric field may excite a soft mode of translational vibrations of $90^0$ walls, which provides considerably larger collective DW mobility than the antiparallel motion of neighboring $90^0$ boundaries. Using a special setup with fingered electrodes, it may be possible to observe a very high dielectric response of an epitaxial film due to appearance of a soft mode of forced DW vibrations.